\documentclass[twocolumn,showpacs,preprintnumbers,amsmath,prd,amssymb,superscriptaddress]{revtex4-1}

\pdfoutput=1

\usepackage{graphicx}
\usepackage{dcolumn}
\usepackage{bm}
\usepackage{xcolor}
\usepackage[unicode=true,pdfusetitle,
 bookmarks=true,bookmarksnumbered=false,bookmarksopen=false,
 breaklinks=false,pdfborder={0 0 1},backref=false,colorlinks=true, citecolor=blue]
 {hyperref}

\def\chiq{\chi_q}
\def\barchiq{\bar{\chi}_q}
\def\mchiq{m_{\chi_q}}

\begin{document}

\title{
  Constraints on millicharged particles with 
  low threshold germanium detectors at 
  Kuo-Sheng Reactor Neutrino Laboratory
}

\author{L.~Singh}
\altaffiliation[Corresponding Author: ]{ lakhwinder@phys.sinica.edu.tw } 
\affiliation{Institute of Physics, Academia Sinica, Taipei 11529, Taiwan}
\affiliation{Department of Physics, Institute of Science, Banaras Hindu University, 
Varanasi 221005, India}

\author{J.W.~Chen}
\affiliation{Department of Physics, Center for Theoretical Physics, and Leung Center for Cosmology and Particle Astrophysics, National Taiwan University, Taipei 10617, Taiwan}
\affiliation{Center for Theoretical Physics, Massachusetts Institute of Technology, Cambridge, MA 02139, USA}

\author{H.C.~Chi}
\affiliation{Department of Physics, National Dong Hwa University, 
Shoufeng, Hualien 97401, Taiwan}

\author{C.-P.~Liu}
\affiliation{Department of Physics, National Dong Hwa University, 
Shoufeng, Hualien 97401, Taiwan}

\author{M.K.~Pandey}
\affiliation{Department of Physics, Center for Theoretical Physics, and Leung Center for Cosmology and Particle Astrophysics, National Taiwan University, Taipei 10617, Taiwan}

\author{H.T.~Wong}
\altaffiliation[Corresponding Author: ]{ htwong@phys.sinica.edu.tw } 
\affiliation{Institute of Physics, Academia Sinica, Taipei 11529, Taiwan}

\author{C.P.~Wu}
\affiliation{Department of Physics, Center for Theoretical Physics, and Leung Center for Cosmology and Particle Astrophysics, National Taiwan University, Taipei 10617, Taiwan}

\author{M.~Agartioglu}
\affiliation{Institute of Physics, Academia Sinica, Taipei 11529, Taiwan}
\affiliation{Department of Physics, National Dong Hwa University, 
Shoufeng, Hualien 97401, Taiwan}
\affiliation{Department of Physics, Dokuz Eyl\"{u}l University, Buca, \.{I}zmir TR35390, Turkey}

\author{M.~Deniz}
\affiliation{Institute of Physics, Academia Sinica, Taipei 11529, Taiwan}
\affiliation{Department of Physics, Dokuz Eyl\"{u}l University, Buca, \.{I}zmir TR35390, Turkey}

\author{H.B.~Li}
\affiliation{Institute of Physics, Academia Sinica, Taipei 11529, Taiwan}

\author{S.T.~Lin}
\affiliation{Institute of Physics, Academia Sinica, Taipei 11529, Taiwan}
\affiliation{College of Physical Science and Technology, Sichuan University, Chengdu 610064, China}

\author{V.~Sharma}
\affiliation{Institute of Physics, Academia Sinica, Taipei 11529, Taiwan}
\affiliation{Department of Physics, Institute of Science, Banaras Hindu University, 
Varanasi 221005, India}

\author{M.K.~Singh}
\affiliation{Institute of Physics, Academia Sinica, Taipei 11529, Taiwan}
\affiliation{Department of Physics, Institute of Science, Banaras Hindu University, 
Varanasi 221005, India}

\author{V.~Singh}
\affiliation{Department of Physics, Institute of Science, Banaras Hindu University, 
Varanasi 221005, India}

\author{Q.~Yue}
\affiliation{Department of Engineering Physics, Tsinghua University, Beijing 100084, China}

\collaboration{TEXONO Collaboration}

\date{\today}

\begin{abstract}
Relativistic millicharged particles ($\chiq$) have been proposed 
in various extensions to the Standard Model of particle physics.
We consider the scenarios where they are produced at nuclear 
reactor core and via interactions of cosmic-rays with the 
earth's atmosphere. Millicharged particles could also be 
candidates for dark matter, and become relativistic through 
acceleration by supernova explosion shock waves. The atomic 
ionization cross section of $\chiq$ with matter are derived 
with the equivalent photon approximation. Smoking-gun signatures 
with significant enhancement in the differential cross section 
are identified. New limits on the mass and charge of $\chiq$ 
are derived, using data taken with a point-contact germanium 
detector with 500~g mass functioning at an energy threshold of 300~eV at 
the Kuo-Sheng Reactor Neutrino Laboratory.

\end{abstract}

\maketitle
\section{Introduction}
The origin of electric charge quantization is one of the 
profound intriguing mysteries of the nature~\cite{Davidson:1991}.
Studies are made to explore the theoretical foundation 
of electric charge quantization in Kaluza-Klein 
higher dimensional theory~\cite{Klein:1926}, the existence of 
magnetic monopoles~\cite{Dirac:1931}, and Grand Unified 
Theories~\cite{Georgi:1974}. Despite consistency of charge 
quantization with all experimental data, the absence of 
evidence for magnetic monopoles and grand unification 
continues to motivate  searches for the existence of non-quantized 
charged particles commonly known as millicharged particles 
and denoted by $\chiq$ with mass $\mchiq$ 
in this article~\cite{Davidson:2000}. Such particles can be 
naturally obtained via including an extra abelian gauge 
U$_{HS}$(1) (the subscript denotes ``Hidden Sector'') into 
the Standard Model (SM) gauge groups~\cite{Holdom:1985, LHC:2015}. 
The SM particles are not charged under this new gauge group, 
while the $\chiq$ under U$_{HS}$(1) acquire small electric charge 
($\delta e_0$) due to the kinetic mixing of SM photon and HS dark 
photon, where $\delta$ is the charge fraction of $\chiq$ and 
$e_0$ is the standard electron charge. The parameter space defined 
by $( \mchiq , \delta )$ is strongly constrained by  
astrophysical observations~\cite{Davidson:1991,Davidson:2000,Mohapatra:1990,Davidson:1994}, 
cosmic microwave background~\cite{Dubovsky:2004,Dolgov:2013}, Big Bang 
nucleosynthesis~\cite{Redondo:2014,Foot:2015bbn}, and direct laboratory 
experiments~\cite{Badertscher:2007,SLAC:1998,CMS:2013}.

The theme of this article is to report new direct laboratory 
limits on $\chiq$ using data acquired by an n-type Point-Contact 
Germanium (PCGe) detector with sub-keV sensitivity 
and excellent energy resolution. The PCGe technology~\cite{Soma:2014} 
is being used in the studies of coherent neutrino nucleus elastic 
scattering, as well as searches of ``light'' 
WIMPs~\cite{TEXONO:2013DM,CDEX:2016prd, CDEX:2018cjp, CDEX:2018prl} 
and neutrinoless double beta decay. 

This article is organized as follows. Three cases of $\chiq$ are 
selected for investigations, whose origins and flux estimations 
are given in Section II. The interaction of $\chiq$ with matter 
and the derivation of cross sections are discussed in Section III.
Physics constraints on $\chiq$ derived from data taken with 
the PCGe at the Kuo-Sheng Reactor Neutrino Laboratory 
(KSNL)~\cite{Soma:2014,Wong:2018, TEXONO:2010nmm, Deniz:2010, TEXONO:2018nDet} are presented in Section IV.

\section{Sources of MilliCharged Particles} 
\label{sect::mQ-source}

\subsection{Nuclear Reactor-$\chiq$}
\label{sect::reactor}
Nuclear reactor cores are powerful sources of $\gamma$-rays 
and therefore have been used for searches of 
axions~\cite{Altmann:1995-Bugey,TEXONO:2007axion},
dark photons~\cite{Park:2017prl,TEXONO:2015DP},  
and millicharged neutrinos~\cite{TEXONO:2014mQ}.
Approximately half of the $\gamma$-radioactivity are 
of ``prompt'' origins directly from highly excited fission 
fragments, while the rest are from
radiative de-excitation of the daughter nuclei, inelastic 
neutron scattering and capture of neutron by core 
materials~\cite{Altmann:1995-Bugey}.
The total prompt $\gamma$-ray spectrum, in the units of MeV$^{-1}$~s$^{-1}$, 
for FRJ-1 (Merlin) reactor core is approximated ( $E_{\gamma}~\geq$~200~keV ) 
by~\cite{Jule-Spez:1984}
\begin{align}
\frac{dN_{\gamma}}{dE_{\gamma}} = 
0.581 \times 10^{18} e^{-1.1 E_{\gamma}({\rm MeV})} \times {\rm Power(MW)}~.
\label{Eq::reactor:flux}
\end{align}
This spectrum has been adopted in earlier
reactor axion~\cite{Altmann:1995-Bugey} and 
dark photon searches~\cite{Park:2017prl}. 

\begin{figure}[hbt]
 \includegraphics[width=8cm]{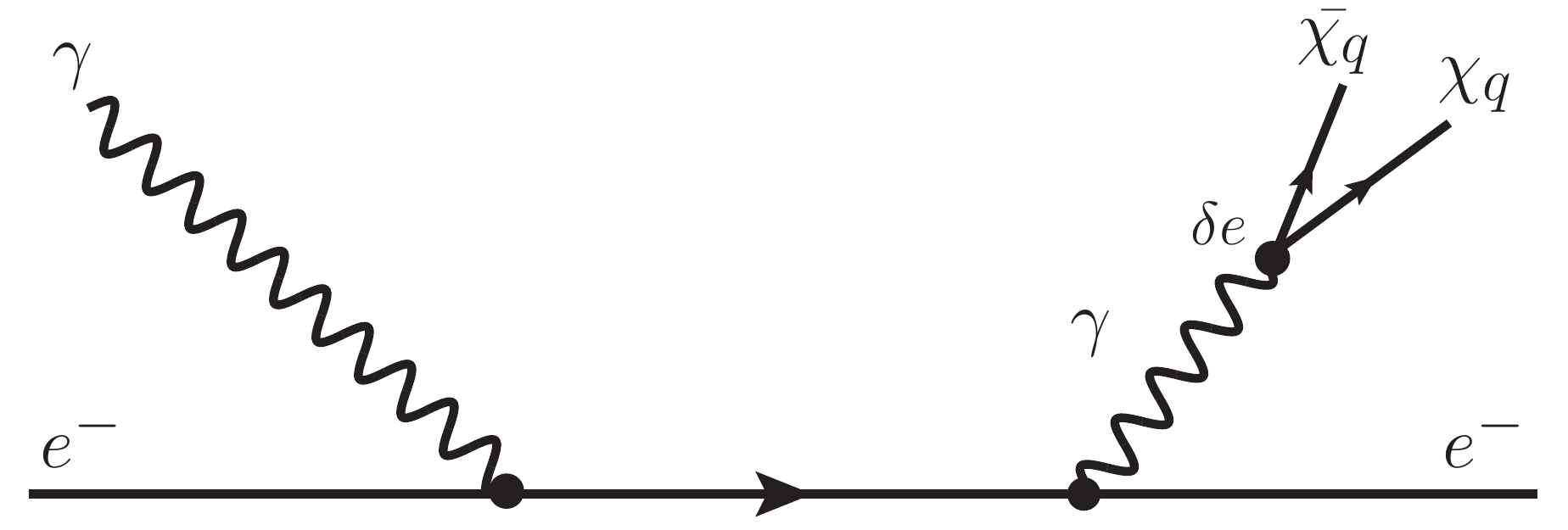}
  \caption{
The production of $\chiq$-$\barchiq$ via Compton-like 
mechanism based on the kinetic mixing of dark photon 
with the SM photon. 
}
 \label{fig::diagram}
 \end{figure}

The light-$\chiq$ can be produced through Compton-like 
processes, where $\gamma$-rays of $\mathcal{O}$(MeV) energy
scatter off electrons in the reactor core
~\cite{Gninenko:2007mQ}. The differential production cross section 
of $\chiq$-$\barchiq$ via a Compton-like process, 
as depicted schematically in Fig.~\ref{fig::diagram},
can be estimated from well known lepton-pair production process 
in the field of an electron. To adapt the lepton-pair production 
formula in the case of $\chiq$ production, 
the $\chiq$-$\bar{\chiq}$ vertex is parameterized
by $\delta e $ and the lepton mass is replaced by $\mchiq$~:

\begin{eqnarray}
\frac{d\sigma}{dE_{\chiq}}
( \gamma e \rightarrow \chiq \barchiq e )
  \simeq  
\frac{4}{3} \frac{\delta^{2}\alpha^{3}}{m_{e}^{2} E^{3}_{\gamma}} 
 ~~ \left[ (3(E_{\chiq}^{2}+E_{\barchiq}^{2}) + \right. \nonumber \\
\left. 2E_{\chiq}E_{\barchiq})  \right] 
{\rm log} \left(
\frac{2E_{\chiq}E_{\barchiq}}{E_{\gamma}m_{\chiq}} \right), 
\label{eq::pp}
\end{eqnarray}
where $m_{e}$ is a mass of the electron; $E_{\chiq}$ and $E_{\gamma}$ are, respectively,
energies of $\chiq$ and $\gamma$-ray, 
while $E_{\barchiq}~=~E_{\gamma}-E_{\chiq}$. 
The differential $\chiq$-flux ($\phi_{\chiq}$) is determined from 
the convolution of reactor $\gamma$-ray spectrum and  
differential production cross section normalized 
by the total cross section ($\sigma_{tot}$),
\begin{align}
   \frac{d\phi_{\chiq}}{dE_{\chiq}} = 
\frac{2}{4 \pi R^{2}}\int \frac{1}{\sigma_{tot}} 
\frac{d\sigma}{dE_{\chiq}} 
\frac{dN_{\gamma}}{dE_{\gamma}}
{dE_{\gamma}} , 
\label{eq::reactorflux}
\end{align} 
where $\it{R}$ is a distance of detector from the center of reactor core.
The factor 2 in Eq.~(\ref{eq::reactorflux}) comes from the 
fact that $\chiq$ particles 
produce in pairs and both can interact at the detectors. 
Although the reactor core produces $\gamma$-rays of  
energy up to 10 MeV, the 1-5~MeV interval of reactor $\gamma$-ray spectrum 
is expected to dominate the $\chiq$-$\barchiq$ pair production 
process of Eq.~(\ref{eq::pp})~\cite{Gninenko:2007mQ}. The Compton scattering 
cross section ($\sigma_{c}$)
of $\gamma$-rays in this energy interval is dominated over atomic 
photoelectric absorption  and pair production cross section, 
even for high Z-materials. Therefore, $\sigma_{tot}$~$\simeq$~$\sigma_{c}$ 
is a reasonable approximation for 1-5~MeV energy interval. A typical reactor core
with 1~GW thermal power produces $\mathcal{O}$ (10$^{20}$) photons per second
at $\mathcal{O}$(MeV) energies. These $\gamma$-rays can interact with electrons 
producing 2.3~$\times~10^{18}\times~\delta^{2}$ number of 
$\chiq$ per second at the reactor core. 

\subsection{Atmospheric-$\chiq$}
\label{sect::atmosphere}
High energy cosmic-rays can produce relativistic-$\chiq$  
when they interact with nucleus in the earth's 
atmosphere~\cite{Dobroliubov:1990mQ,Perl:2009review}.
Bremsstrahlung  and multiperipheral (which include the 
interaction of two virtual photons) are  main production mechanisms 
of $\chiq$ in the collision of 
charged particles and their cross sections are proportional to  
$\delta^{2} \alpha^{4} Z^{2}$ and $\delta^{4} \alpha^{4} Z^{2}$,
respectively~\cite{Dobroliubov:1990mQ}.
 The energetic cosmic-rays may produce $\chiq$ with masses 
inaccessible to current particle accelerators.

The investigation of cosmic-rays can provide constraints 
on $\chiq$ via interactions between $\chiq$ and detectors. 
The energy loss of $\chiq$ through excitation and ionization 
is proportional to $\delta^{2}$, which is much lower  
than the minimum ionizing particles of unit charge 
under the similar conditions. These exotic $\chiq$ are 
therefore also called Lightly Ionizing Particles and 
fractional charged particles in the 
literature~\cite{Perl:2009review,Napolitano:1982,Hass:2015,Cairns:1969}. 
Since these particles deposit little energy at detectors, 
the high resolution and sub-keV threshold of PCGe 
make it an ideal detector for their searches.

The mass range of cosmogenic produced $\chiq$ is unknown 
due to unknown production conditions. Therefore, the 
experimental sensitivity is usually expressed in terms of the 
integral incoming flux (I$_{\chiq}$) in the units of cm$^{-2}$ s$^{-1}$ sr$^{-1}$ 
as a function of $\delta$~\cite{Perl:2009review}. 

\subsection{Dark Cosmic-Rays $-$ DM-$\chiq$}
\label{sect::darkmatter}

If self-interaction between dark matter (DM)
is mediated by massless dark photon, the gauge invariance 
mandates that DM  would be multi-component~\cite{Petraki:2014}. 
Two oppositely charged massive 
fermion interacting through a massless 
U(1)$_{HS}$ gauge boson would give rise 
to bound states which, implies atomic DM~\cite{Cline:2012}. 
The atomic dark sector also requires 
the existence of a dark matter-antimatter asymmetry 
in the early universe to set the relic abundance of DM.
The self-interaction of dark matter is a interesting
scenario which has the potential to explain the observed
properties of galaxies small-scale as well as
large-scale structures~\cite{Foot:2016jcap,FOOT:2015plb}. 
The dark sector particles may have both neutral and 
ionized components. The latter component may be due 
to incomplete recombination of primordial 
DM gas~\cite{Francis:2013} and 
re-ionization by sources such as starlight and supernova 
explosions~\cite{Foot:2014}  which can efficiently
overcome the binding energy of dark atoms. 
The ionized components of DM could be accelerated analogous to 
standard cosmic-rays~\cite{PingHu:2017}.

The diffuse-shock acceleration is widely accepted as being
responsible for accelerating the standard cosmic-rays
at the blast waves of supernova remnant~\cite{Blandford:1978ApJ}.
Ambient charged particles captured by supernova shock wave 
fronts can be accelerated to high energies during the 
lifetime of a supernova remnant. The key feature of this 
theory~\cite{Blandford:1987PR}, usually known as first order Fermi 
acceleration, is that the fractional energy gain by charged 
particles via each crossing of supersonic shock front and 
scattered back from turbulent magnetic fields associated 
with the shock front is proportional to the first order of shock velocity.
This mechanism also naturally explains the commonly observed power 
law cosmic-rays energy spectrum. 
The maximum energy attained by 
$\chiq$ of charge fraction $\delta$ 
is simply product of 
rate of energy gain and time spent in shock. 
The upper limit of maximum energy can be obtained by assuming a 
minimum diffusion length equal to gyroradius of $\chiq$
\begin{equation}
 E_{max} ~ \simeq ~ \delta e_0 ~ B ~ t_{A} ~ V_{S}^{2}. 
\end{equation}
A typical supernova can accelerate a proton to 
$10^{6}$~GeV energy level with the following parameters such as 
the shock wave speed $V_{S}\sim 10^{4}~{\rm km~s^{-1}}$, 
total acceleration time $t_{A} \approx 10^{3}$ 
years, and magnetic field $B \sim 10^{-10}$~T. 
If the gyroradius of the $\chiq$ as DM candidates
is smaller than the length of shock waves, 
they can be efficiently accelerated by the Fermi mechanism. 
Therefore, accelerated DM-$\chiq$ may accompany ordinary 
cosmic-rays arriving on the earth.

The precise prediction of the DM-$\chiq$
``dark cosmic-ray'' flux requires knowledge of 
the injection spectrum. An estimation is obtained 
in Ref.~\cite{PingHu:2017} by using results of 
ion acceleration in shocks. 
The energy spectrum in the units of ${\rm cm^{-2}~s^{-2}~sr^{-1}~GeV^{-1}}$,
follows a simple power law: 
\begin{equation}
\frac{d \phi_{\chiq}}{dE_{\chiq}} ~ =  ~
30 ~ \delta^{\alpha-1} ~
\left(\frac{\rm GeV}{m_{\chiq}}\right) 
\left(\frac{E_{\chiq}}{\rm GeV}\right)^{-\alpha}, 
\label{eq::dcrflux}
\end{equation}
where $\alpha$ = 2.7 is the differential spectral index 
of the ordinary cosmic-ray flux.  
The maximal attainable energy of DM-$\chiq$
would be suppressed by a factor of $\delta$ 
from the maximum energy of the cosmic-ray proton.

\section{Atomic Ionization with Millicharged Particles}
\label{sect::mQ-diffcs}

\begin{figure}[hbt]
\begin{center}
{\bf (a)}\\
 \includegraphics[width=8cm]{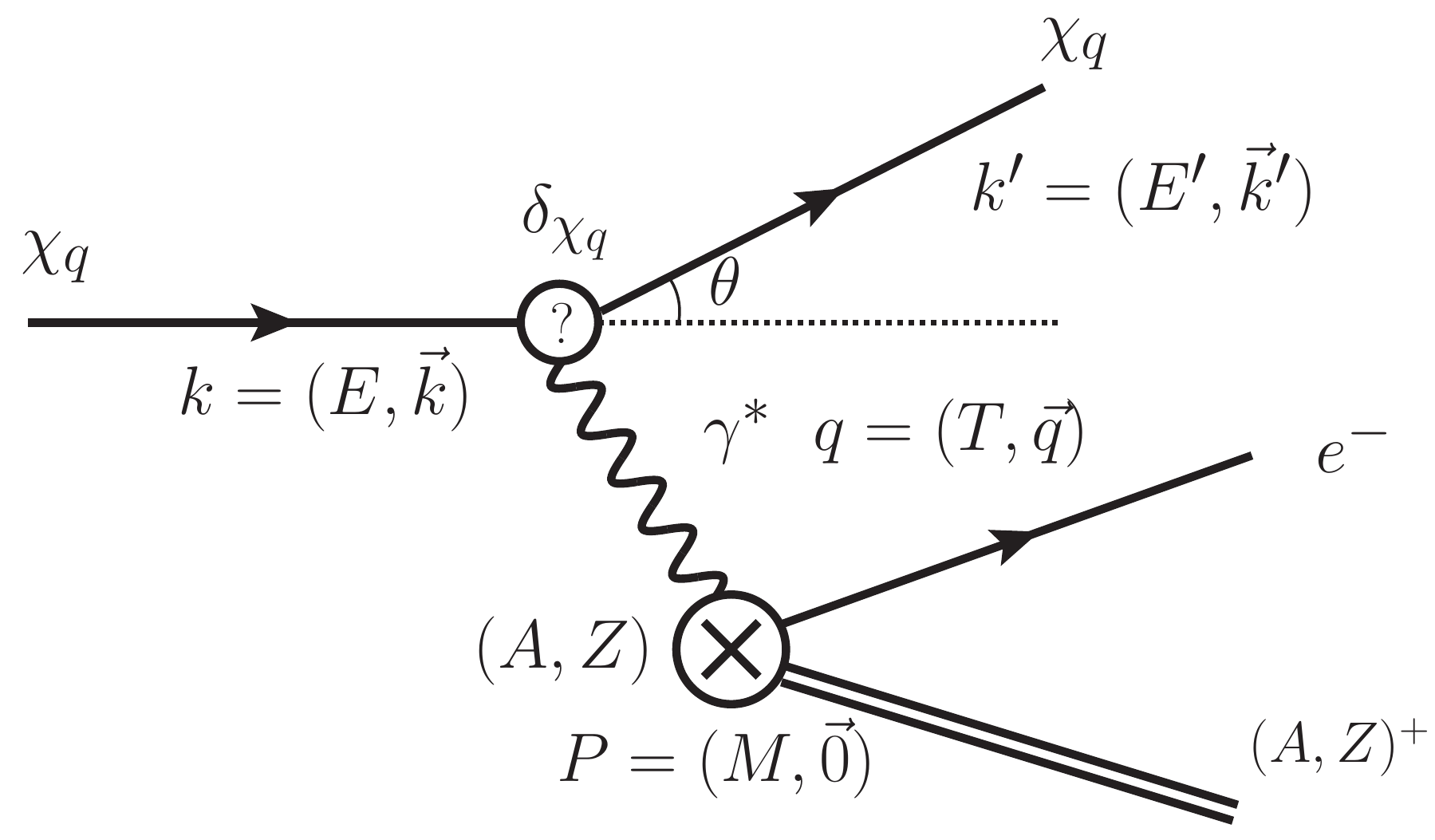}\\
{\bf (b)}\\
 \includegraphics[width=8.5cm]{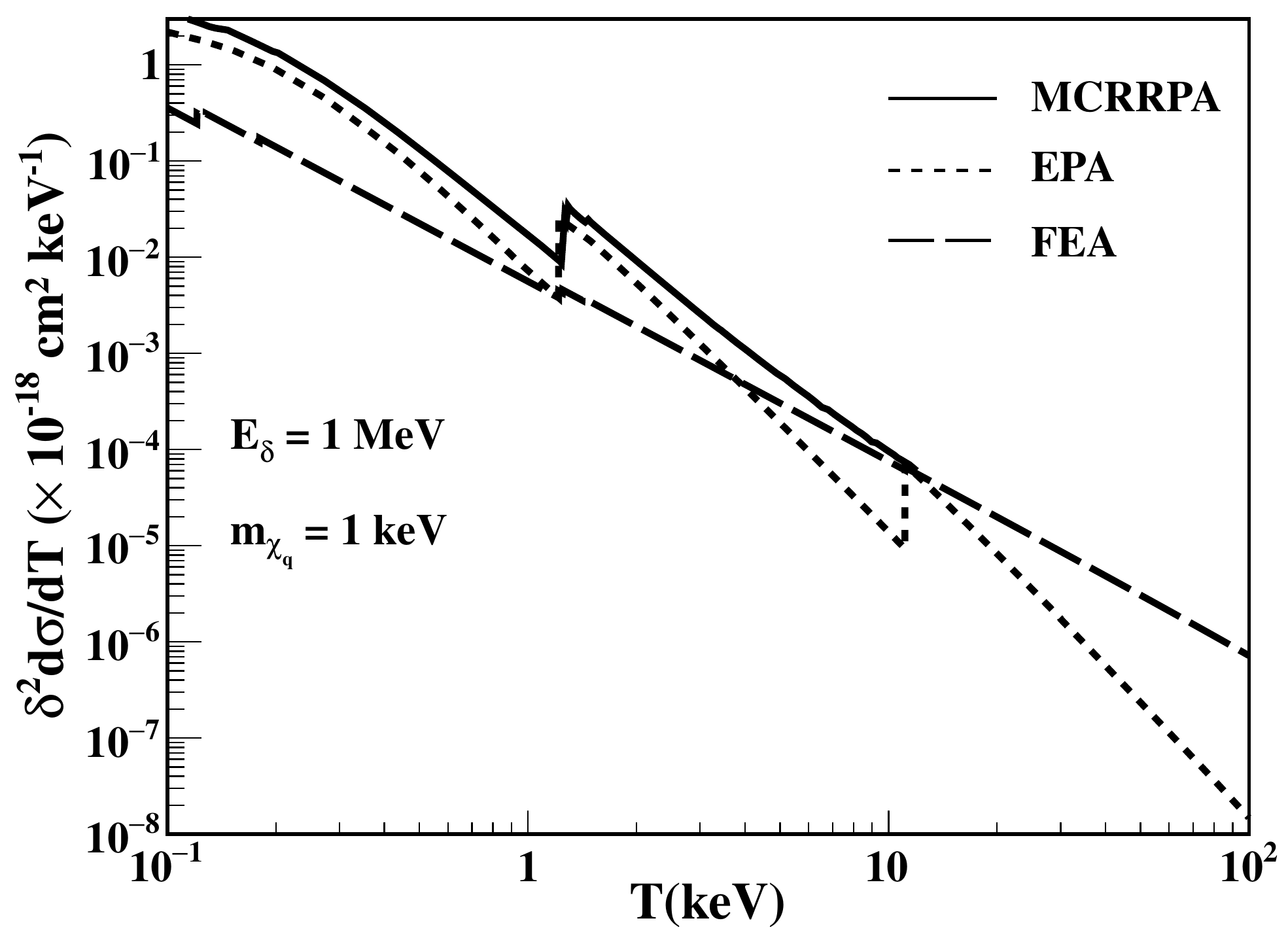}
  \caption{
(a)
Schematic diagram of a relativistic-$\chiq$ with charge fraction $\delta$ interaction 
via atomic ionization channel. 
(b)
The differential scattering cross sections of Ge-ionization by $\chiq$ 
with $\mchiq$ = 1 keV, $\delta$ = 1 and monochromatic $E_{\chiq}$ = 1 MeV 
are derived for FEA (solid line) 
and EPA (dashed line).
}
\label{fig::chiAI}
\end{center}
\end{figure}

The millicharged $\chiq$'s are relativistic
and can interact electromagnetically with matter 
via atomic ionization 
\begin{equation}
\chiq ~ + ~  A ~ \rightarrow ~ \chiq ~ + ~ A^+ ~ + ~ e^- ,
\label{eq::chiAI}
\end{equation}
through the t-channel process depicted in
Fig.~\ref{fig::chiAI}a. The cross sections 
for the analogous atomic ionization processes due to
neutrino electromagnetic interactions were recently 
derived~\cite{TEXONO:2014mQ,JWChen:2015nep}. The case of dark matter 
sterile neutrino as a non-relativistic probe was also 
investigated~\cite{JWChen:2016sn}. This work expands these 
studies to one with having a relativistic probe with finite mass. 

The differential cross section with respect to the energy transfer ($T$)
can be expressed as 
\begin{equation}
\frac{d\sigma}{dT} = 
\delta^{2}\int \left[ \frac{2\pi\alpha^{2}}{q^{4}} 
\frac{|\vec{k_{2}}|}{|\vec{k_{1}}|} (V_{L}R_{L} + V_{T}R_{T}) \right]
d ( cos\theta ) ,
\label{eq::dsigmadT}
\end{equation}
where $\vec{k_{1}}$ and $\vec{k_{2}}$ are, respectively, the 3-momenta 
of the incoming and outgoing $\chiq$,
$q^2$ is the squared 4-momentum transfer,  
$R_{L}$ and $R_{T}$ are the 
longitudinal and transverse response functions, 
while $V_{L}$ and $V_{T}$ are the 
longitudinal and transverse kinematic factors 
for electromagnetic interaction. 
The response functions can, in principle, be calculated by 
many-body wave functions. 
In practice, for most cases, 
the calculations are highly non-trivial 
and schemes such as 
free electron approximation (FEA) and equivalent 
photon approximation (EPA) provide good estimations
at certain kinematic regions.

The cross section evaluation with FEA 
assumes electrons occupying atomic orbitals 
as individual and independent.
The electrons are inactive 
when energy transfer is below the ionization threshold for 
their corresponding orbitals. 
Sharp step functions are therefore produced in 
the cross section. 
Derivations with FEA
take the specific kinematics case $q^{2} = -2 m_{e} T$ 
(where {\it{T}} is the energy transfer) 
fixed by energy-momentum conservation. 
For high-energy scattering where the atomic binding energy 
is negligible, FEA is a good approximation.
The differential cross section for $\chiq$ elastic scattering 
off electron is given by
\begin{eqnarray}
\left. \frac{d\sigma}{dT} \right| _{\rm FEA} & = & 
\frac{\delta^{2} \pi \alpha^{2}}
{(m_{e}^{2}T^{2})(E^{2}_{\chiq}-m^{2}_{\chiq})}  \\
 & & 
\left[ m_{e}(E^{2}_{\chiq}+(E_{\chiq}-T)^{2})
-T(m^{2}_{e}+m^{2}_{\chiq}) \right] . \nonumber
\end{eqnarray}
However, FEA largely underestimates the scattering cross section
at $T$ comparable to the atomic transition energy scale~\cite{JWChen:2013Hyd}.

In kinematical region, where the momentum transfer 
by highly relativistic charged particles 
is small, EPA is a good approximation in the cross section calculations.
In the limit of $q^{2}\rightarrow~0$, the contribution from the longitudinal 
polarized virtual photons vanishes and  that from transversely polarized 
virtual photons approaches the cross section due to real photons. 
Furthermore, the on-shell transverse response function 
is directly related to
the total photo-absorption cross section
\begin{equation}
\sigma^{\gamma}_{abs}(T) ~ = ~ \frac{2\pi^{2}\alpha}{T} ~ R^{\gamma}_{T}(q^{2}=0).
\end{equation}
As a result, the differential cross section of~Eq.~(\ref{eq::dsigmadT}) 
can be approximated by 
\begin{eqnarray}
\left.\frac{d\sigma}{dT}\right|_{\rm EPA} & = &
\delta^{2}\int dcos\theta \frac{2\pi\alpha^{2}}{q^{4}} 
\frac{|\vec{k_{2}}|}{|\vec{k_{1}}|} \left[V_{L}
\left(\frac{T}{2\pi^{2}\alpha}\sigma^{\gamma}_{abs}(T)\right)\right] \nonumber \\ 
& = & \delta^{2}\frac{\alpha}{\pi}
\frac{|\vec{k_{2}}|}{|\vec{k_{1}|}} T ~ \sigma_{\gamma}(T)
\int \left[  \frac{V_{T}}{q^{4}} \right] d ( cos \theta ) ,
\end{eqnarray}
 where the transverse part for Coulomb interaction is
described by 
\begin{equation}
V_{T} = - \left[2m^{2}_{\chiq} + q^{2} + 
\frac{q^{2}(q^{2}+4E_{\chiq}(E_{\chiq}-T))}{2|\vec{q}~|^{2}} \right],
\end{equation}
and the scattering angle $\theta$ depends on $\vec{k_{1}}$,  $\vec{k_{2}}$ and $\vec{q}$. 
Upon integration after
a change of variable from $d ( cos\theta )$ to $d {\vec{q}}^{~2}$, 
the EPA cross section for ultra-relativistic particle can be written as
\begin{equation}
\left. \frac{d \sigma}{dT} \right| _{\rm EPA} ~ =  ~ 
\delta^2 \left[ \frac{2 \alpha_{em}^2}{\pi} \right]
\left[ \frac{\sigma_{\gamma}(T)}{T} \right]
{\rm log} \left[ \frac{E_{\chiq}}{m_{\chiq}} \right].
\label{eq::dcsEPA}
\end{equation}

The derived differential cross sections of $\mchiq$=1~keV at $E_{\chiq}$=1~MeV 
on Ge target under FEA and EPA schemes are depicted in Fig.~\ref{fig::chiAI}b.
Although the EPA is a good approximation in the most interesting 
sub-keV region of $\it{T}$, it underestimates the scattering cross section above 
a few keV region of $\it{T}$ where FEA works well. The EPA and FEA schemes 
therefore serve as conservative approximations in the region near and away from 
ionization thresholds, respectively. 
Accordingly, the scattering cross sections are combined in this analysis 
such as
\begin{equation}
  \left. \frac{d \sigma} {dT} \right|_{\rm tot}~=~max\left (\left. \frac{d \sigma}{dT}\right|_{\rm EPA},~\left. \frac{d \sigma}{dT}\right|_{\rm FEA}\right ).
  \label{EQ:combdcs}
\end{equation}

The uncertainties of the differential cross section are estimated via comparisons with
an $ab$-$initio$ many-body calculation using multi-configuration relativistic random-phase
approximation (MCRRPA) which has been successfully applied to photoexcitation,
photoionization, and neutrino-induced ionization of germanium and xenon atomic
systems~\cite{JWChen:2013plb,TEXONO:2014mQ,JWChen:2017plb}.
As illustrated in~Fig.\ref{fig::chiAI}b the
combined cross sections of~Eq.~(\ref{EQ:combdcs}) from EPA and FEA is
30\% less than those due to MCRRPA calculations at energy away from the
discrete excitation thresholds. This scheme will give conservative estimate
in deriving limits. The ``smoking gun'' experimental signatures of 
$\chiq$ are observation of K- and L-shell peaks at the specific binding energies 
in the energy spectra, with known intensity ratios. 
This feature requires excellent energy resolution of 
the PCGe detectors where keV X-ray peaks can be resolved.

\section{EXPERIMENTAL CONSTRAINTS}

Analysis results presented in this article are
based on data taken with an n-type PCGe at KSNL.   
The facility~\cite{Wong:2018, TEXONO:2010nmm, Deniz:2010, TEXONO:2018nDet} is located at a distance of 28~m from a 2.9~GW 
thermal power nuclear reactor core with 30~meter-water-equivalent  
overburden. The n-type PCGe is enclosed by a NaI(Tl) anti-Compton 
(AC) detector of mass 38.3 kg. This setup is placed inside the 
50~ton shielding structure equipped with  cosmic-ray (CR) 
veto scintillator panels. 
Data with  n-type PCGe are selected for its low-threshold at 300~eV, 
and the absence of 
anomalous surface events which serves to minimize complications to the
analysis~\cite{Soma:2014}.
A total of 124.2/70.3~kg-days of reactor ON/OFF data
are adopted for this analysis. 

\begin{figure}[hbt]
{\bf (a)}\\
\includegraphics[width=8.5cm]{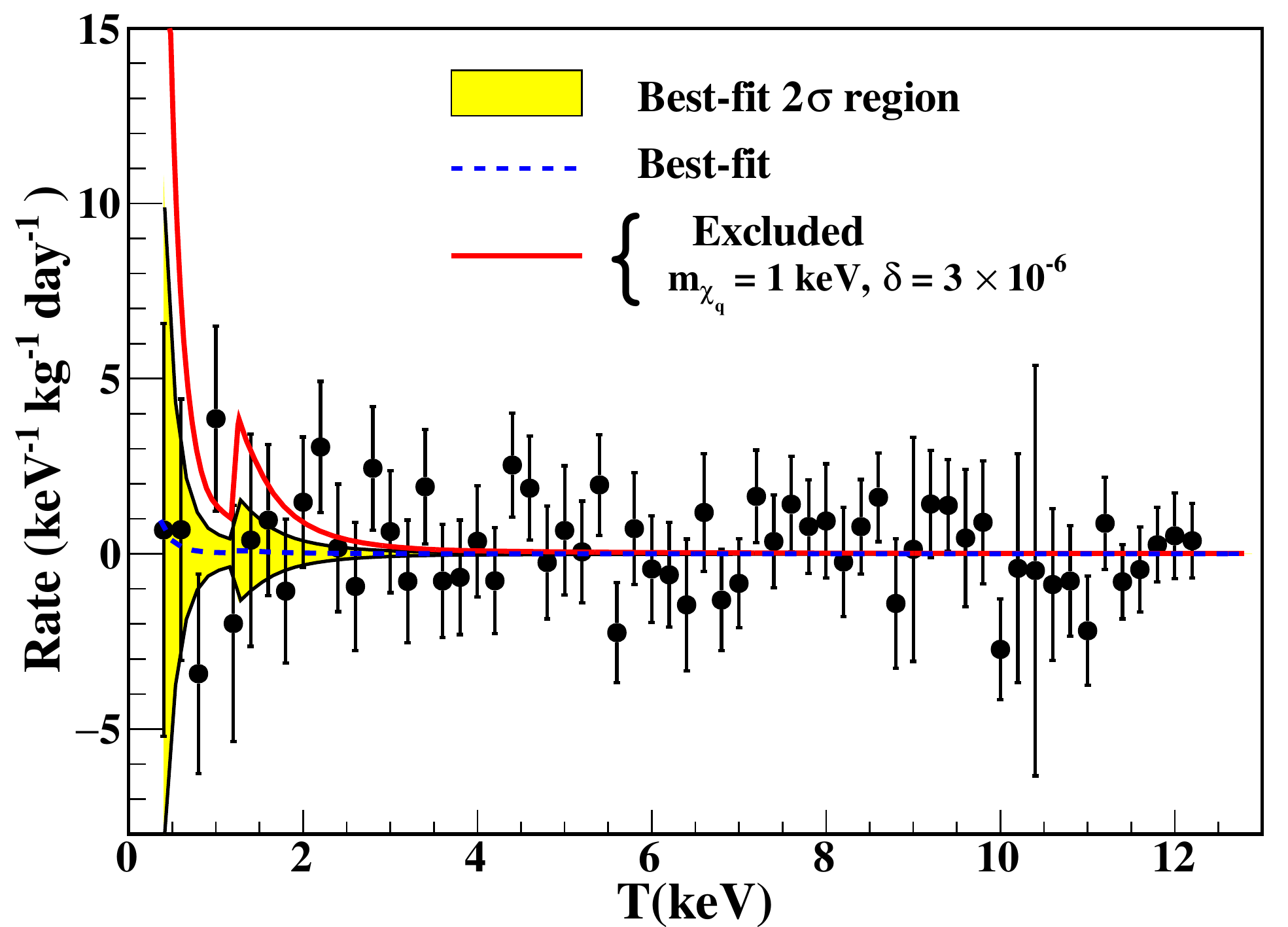}\\
{\bf (b)}\\
\includegraphics[width=8.5cm]{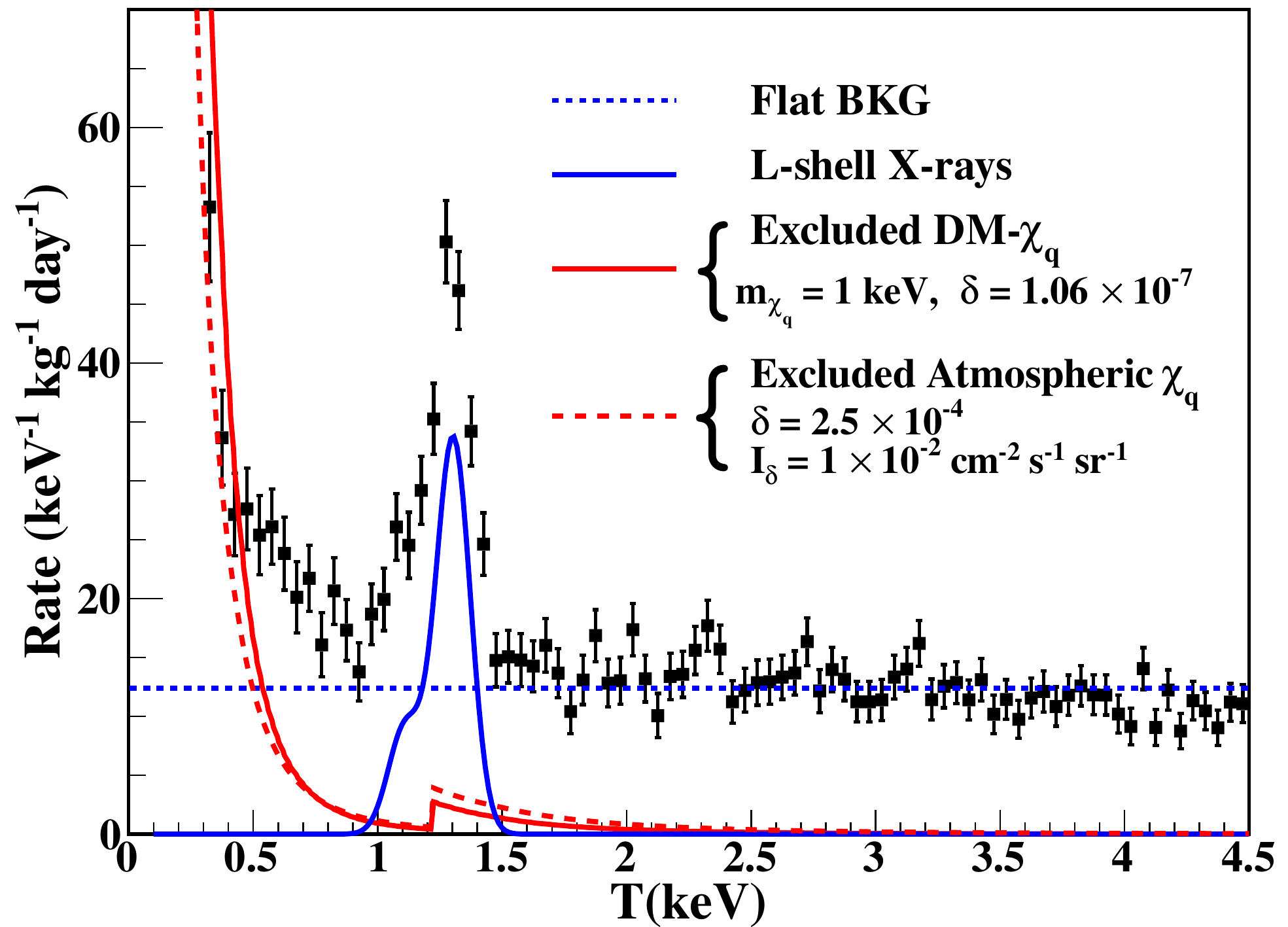}
\caption{
(a)
Reactor ON$-$OFF residual spectrum of n-type PCGe
with AC$^{-}$ $\otimes$ CR$^{-}$ selection.
The best-fit 2$\sigma$ region of possible 
$\chiq$ interactions are shown as 
the yellow band, with an excluded scenario
of reactor-associated $\chiq$ at specified 
($\mchiq$~=~1~keV, $\delta$~=~3~$\times$~10$^{-6}$) 
is superimposed.
(b)
The total AC$^{-}$ $\otimes$ CR$^{-}$ spectrum
showing a flat background due to ambient
high-energy $\gamma$-rays and the L-shell X-rays
from internal radioactivity.
Excluded scenarios of atmospheric  and DM-$\chiq$ at specified ($\delta$~=~2.5~$\times$~$10^{4}$, I$_{\delta}$~=~1~$\times$~10$^{-2}$~cm$^{-2}$s$^{-1}$sr$^{-1}$) and ($\mchiq$~=~1~keV,  $\delta$~=~1.06~$\times$~10$^{-7}$), respectively, 
are superimposed. }
\label{fig::npc-spec}
\end{figure}

\begin{figure}[!hbt]
{\bf (a)}\\
  \includegraphics[width=8.0cm]{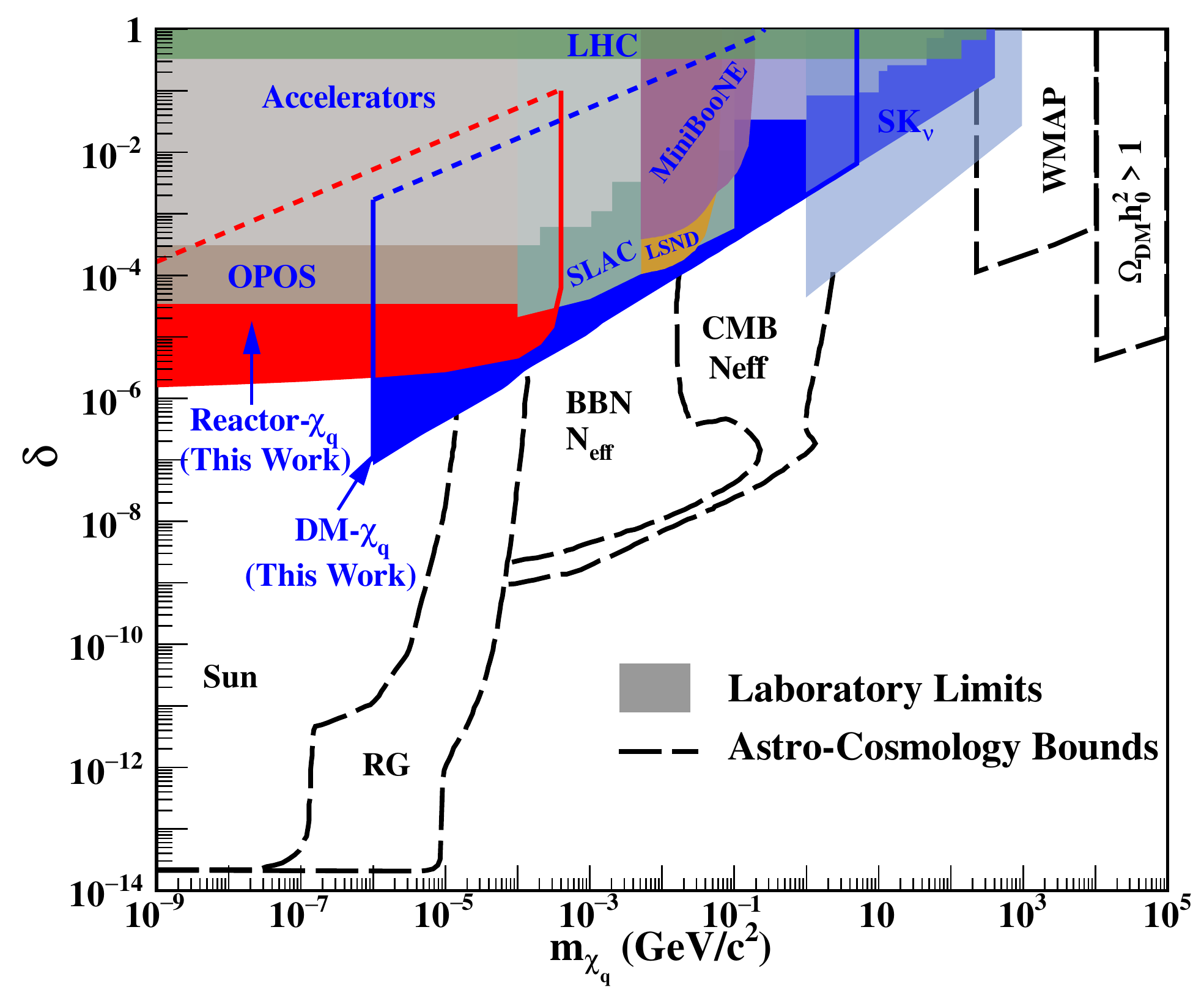}\\
{\bf (b)}\\
\includegraphics[width=8.0cm]{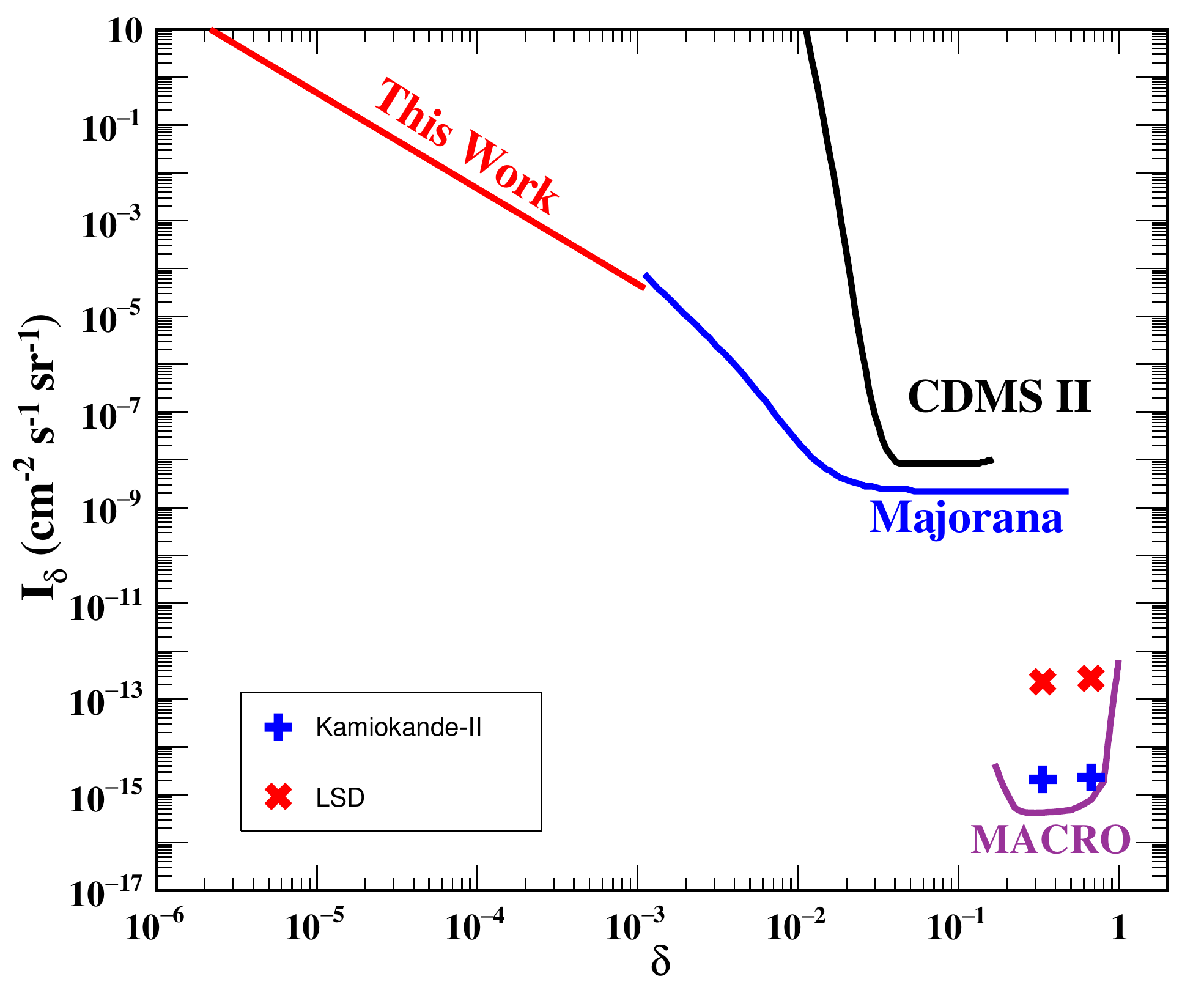}
\caption{
(a) 
Exclusion regions at 90\% C.L. in ($\mchiq$, $\delta$) parameter space
for millicharged particles with a massless dark photon. 
Cosmological and astrophysical bounds from Ref.~\cite{Vinyoles:2016jcap}
are denoted as dotted lines. The direct laboratory limits  
from Accelerators~\cite{Davidson:2000}, OPOS~\cite{Badertscher:2007},
SLAC~\cite{SLAC:1998}, LHC~\cite{CMS:2013}, LSND~\cite{Magill:2018arXiv}, MiniBooNE~\cite{Magill:2018arXiv} and SK-$\nu$~\cite{PingHu:2017} experiments are 
represented as shaded regions. 
New 90\% C.L. excluded regions of this work with $\chiq$ from reactor
and dark primary cosmic-rays are shown as red and blue shaded areas,
respectively. The dotted lines correspond to the upper bounds of the
exclusion regions, due to complete attenuation of $\chiq$ before reaching the detector. 
(b) 
Excluded  parameter space at 90\% C.L. on incoming flux of $\chiq$
from secondary dark cosmic-rays versus its charge fraction $\delta$.
Results from other benchmark experiments such as  LSD~\cite{Aglietta:1994AP}, 
Kamiokande-II~\cite{Mori:1991prd}, MACRO~\cite{Ambrosio:2004arXiv},CDMS II~\cite{Agnese:2015prl},  
and MAJORANA~\cite{Alvis:2018prl} are superimposed. The constraints are applicable
to $\mchiq~>~4 \times 10^{-4}$~GeV/c$^{2}$, below which attenuation effects before reaching the detectors would
limit the sensitivities.
}
\label{fig::ex-plot}
\end{figure}


Every Ge-trigger is categorized 
by AC$^{+(-)}$ $\otimes$ CR$^{+(-)}$, where the superscript -(+) 
denotes anti-coincidence (coincidence) of the Ge-signals
with the AC and CR detectors~\cite{TEXONO:2013DM}. 
The AC$^{-}$ $\otimes$ CR$^{-}$ events are uncorrelated 
with other active veto detector systems 
and are therefore candidates of neutrino, WIMP and other exotic events.
The Reactor ON$-$OFF residual spectrum and the raw rates for
the AC$^{-}$ $\otimes$ CR$^{-}$ events are depicted in Fig.~\ref{fig::npc-spec}a
and Fig.~\ref{fig::npc-spec}b, respectively.

The expected differential count rates due to possible $\chiq$ interaction
with matter are obtained by integrating the
$\chiq$-flux formulae of Section~\ref{sect::mQ-source}
with the differential cross sections 
of Section~\ref{sect::mQ-diffcs}:
\begin{equation}
\frac{d R}{d T} ~ = ~ \rho_{A} 
 \int_{E_{min}}^{E_{max}}
\left[ \frac{d\sigma}{dT} \right] ~ 
\left[ \frac{d\phi_{\chiq}}{dE_{\chiq}} \right] ~
dE_{\chiq}  ,
\label{eq::exptrate} 
\end{equation}
where $\rho_{A}$ is atomic number density per unit target mass
and $( E_{min} , E_{max} )$ are the (minimum, maximum) energy
of $\chiq$.
Constraints from each of the three discussed $\chiq$ channels
are derived from the measured AC$^{-}$ $\otimes$ CR$^{-}$ spectra  
after subtraction of (i) internal radioactivity 
due to K/L-shell X-rays from cosmogenically-activated
isotopes in the Ge-target, and (ii) a flat background estimated
from ambient high-energy $\gamma$-rays, following background 
understanding and analysis procedures from earlier work on 
similar detectors and configurations~\cite{TEXONO:2013DM,CDEX:2016prd, CDEX:2018cjp, CDEX:2018prl}.

The attenuation of relativistic-$\chiq$ between source and detector is
estimated by via the radiation length~($X_{0}^{e}$) of high energy electrons
pre-dominantly through bremsstrahlung in matter. The analogous radiation
length~($X_{0}^{\chi}$) of the relativistic-$\chiq$ is given by 
\begin{equation}
    \frac{X_{0}^{\chi}}{X_{0}^{e}} \simeq \frac{1}{\delta^{4}} \left(\frac{m_{\chi}}{m_{e}}\right)^{2} ,
    \label{EQ:Ecut-off}
\end{equation}
where $m_{e}$ is mass of the electron. The KSNL reactor source is approximated
by 10~m of water ($X_{0}^{e}$~=~36.08~cm), 10 m of concrete ($X_{0}^{e}$~=~11.55~cm)
and 50 cm of lead ($X_{0}^{e}$~=~0.55~cm), while the atmospheric source is given
by 20 km of air ($X_{0}^{e}$~=~29.89), 10 cm of concrete and 50 cm of lead. The
attenuation effects place upper bounds to the excluded regions in the analysis.

\subsubsection{Reactor-$\chiq$}
The reactor ON/OFF comparison 
provides a sensitive probe for laboratory searches on exotics 
particles. A positive signal of reactor-$\chiq$ discussed in 
Section II-A would manifest itself 
as excess of events with 1/$T^{2}$ profile in the ON$-$OFF 
``residual''  energy spectra. The  kinematic threshold of $\chiq$ 
pair production in the field of electron is 
2$\mchiq$($\mchiq$ + $m_{e}$)/$m_{e}$, therefore the search of 
reactor-$\chiq$  is only sensitive 
to $\mchiq$ $<$ 1 MeV. The millicharged $\chiq$ produced by 1 to 5~MeV 
$\gamma$-rays are found to be broadly distributed over 
the energy range 0 $<$ $E_{\chiq}$~$<$~5~MeV, with a few percent 
below 50~keV.

More than 50\% of the $\gamma$-fluxes of nuclear reactor cores are due to
prompt-$\gamma$ of highly excited fission fragments, with slight dependence
on the actual fuel composition~\cite{Altmann:1995-Bugey}. This introduces an uncertainty
to prompt $\gamma$-flux Eq. 1, giving a 10\% systematic error to reactor-$\chiq$ fluxes.

The expected rate of $\chiq$ producing in reactor at Ge-detector, 
in the units of kg$^{-1}$ keV$^{-1}$ day$^{-1}$, is derived from  
Eq.~(\ref{eq::exptrate}) by using flux from Eq.~(\ref{eq::reactorflux}).  
To extract upper limits on $\delta$ as a function of $\mchiq$, 
the  minimum-$\chi^{2}$ fit method is adopted. Systematic uncertainty in flux ($\pm$10\%)
is taken into account. The input are varied within one-RMS range, and the most conservative
constraints are adopted.
The excluded ($\mchiq$, $\delta$) 
parameter space at 90\% C.L. is thus derived and displayed in 
Fig.~\ref{fig::ex-plot}a as a solid deep-red region.  
Previous laboratory limits~\cite{Davidson:2000, Badertscher:2007, SLAC:1998,PingHu:2017, CMS:2013} as well as bounds from 
cosmological and astrophysical model~\cite{Vinyoles:2016jcap}, are superimposed. 
The residual spectrum with best-fit for $\mchiq$~=~1 keV is 
depicted in Fig.~\ref{fig::npc-spec}a and the 2$\sigma$ 
uncertainty band is superimposed.  
These new results provide improved limits at light-$\chiq$ of  
$\mchiq$~$<$~4$\times$10$^{-4}$~GeV/c$^{2}$.

\subsubsection{Atmospheric-$\chiq$}
The high energy cosmic-rays are capable of creating massive $\chiq$ 
via interaction with the earth's atmosphere, as discussed in
Section II-B. In this analysis, we focus on $\chiq$ with 
$\delta$ $<$ $10^{-3}$ and the signal would be neutrino-like, 
where $\chiq$ interact only once with the Ge-target, 
without producing signals at CR and AC veto detectors. 

The upper limits on I$_{\chiq}$ are derived using
the most conservative assumption that predicted signal 
cannot exceed the measured background  AC$^{-}$ $\otimes$ CR$^{-}$ 
rate depicted in Fig.~\ref{fig::npc-spec}b. 
The differential cross section in Eq.~(\ref{eq::dcsEPA})
is independent of $E_{\chiq}$ for ultra-relativistic condition.
The theoretically expected spectrum of $\chiq$-induced 
events is
\begin{equation}
\frac{d R}{d T} ~ = ~ \rho_{A}\phi_{\chiq} \frac{d\sigma}{dT} ,
\label{eq::expt-rate-reactor}
\end{equation}
where $\phi_{\chiq}$ is the only free parameter 
used to fit the measured  spectrum.
The corresponding upper limits on I$_{\chiq}$ as a function 
of $\delta$ are displayed in Fig.~\ref{fig::ex-plot}b and 
constraints from previous experiments~\cite{Aglietta:1994AP,Mori:1991prd, Ambrosio:2004arXiv,Agnese:2015prl, Alvis:2018prl} 
are also superimposed. The constraints are applicable to $\mchiq~>~4 \times 10^{-4}$~GeV/c$^{2}$,
below which attenuation of $\chiq$-fluxes by radiation effects before
reaching the detector would limit the sensitivities.

The differential cross section due to $\chiq$-interaction 
has enhancement in sub-keV region of $\it{T}$ as shown in Fig.~\ref{fig::chiAI}b, 
thereby extending the lower reach of $\delta$ from 10$^{-3}$ to new 
domain of 10$^{-6}$. An excluded scenario of ($\delta$~=~2.5~$\times$~10$^{-4}$, 
I$_{\delta}$~=~1~$\times$~10$^{-2}$ ~cm$^{-2}$~s$^{-1}$~sr$^{-1}$)  
is shown in Fig.~\ref{fig::npc-spec}b as illustration.

\subsubsection{Dark Cosmic-Ray DM-$\chiq$}
As discussed in Section II-C, dark matter ion can be accelerated 
by supernova shock waves. In the absence of injection
spectrum, the prediction of dark cosmic-ray flux is strongly model-dependent,
in which a conservative systematic uncertainty of $\pm$50\% is assigned. Subsequent
analysis follows the same procedures as described in the reactor-$\chiq$ case. 
The expected yield of the DM-$\chiq$ 
events in the detector is calculated by mean of 
Eq.~(\ref{eq::exptrate}) adopting flux from Eq.~(\ref{eq::dcrflux}).
The minimum energy ($E_{min}$) of dark cosmic-ray 
is adopted with the assumption of their being ultra-relativistic, 
or $E_{min}>\gamma \mchiq$. The value of $\gamma$ 
is assigned to be 10, where EPA and exact calculation for cross section 
are consistent~\cite{Hencken:1996prc}. The maximum energy of 
dark cosmic-ray is taken as $E_{max}$~=~$\delta \times $10$^{6}$~GeV.

The evaluation of upper limits for various mass follows the conservative 
assumption that the predicted signal cannot exceed the measured rates. 
For fixed $\mchiq$, $\delta$ is the only free parameter derived by 
minimum-$\chi^{2}$ fit method with a flat background assumption.  
The model-dependent excluded regions on $\delta$ at 90\% C.L. for $\mchiq$
between 10$^{-6}$~GeV/c$^{2}$ to 1~GeV/c$^{2}$ are depicted in Fig.~\ref{fig::ex-plot}a as
shaded blue region. We note that the bounds are strongly dependent on the DM-$\chiq$ model
adopted and therefore not at comparable solid footing as those for reactor-$\chiq$. The blue dotted-lines denote boundaries of exclusion
due to the attenuation of relativistic-$\chiq$ between source
and detector. The lower reach of $\mchiq$ is defined by $E_{min} > \gamma~\mchiq$, the minimum
energy required by DM-$\chiq$ to produce recoil electron above the minimum detectable energy. 
Example of an excluded spectrum is shown in Fig.~\ref{fig::npc-spec}b 
for $\mchiq$~=~1~keV and $\delta$~=~1.06~$\times$~10$^{-7}$.

\section{Conclusion and Prospects}
Hidden sector with massless gauge boson allows possibilities 
of multicomponent dark matter. Its ionic constituents can acquire 
small charges (millicharge) under the hidden sector gauge group.  
In this paper, we have derived the direct constraints on $\chiq$
with low threshold point-contact germanium detectors under the 
scenarios of $\chiq$  produced at (i)~nuclear power reactors, 
(ii)~as products of cosmic-rays interactions, and (iii)~as dark 
matter particle accelerated by supernova shock.  The sub-keV sensitivity 
leads to improve direct limits of $\delta$ at small $\mchiq$ and extend 
the lower reach of $\delta$ to 10$^{-6}$.

\section{Acknowledgments}
This work is supported by the Academia Sinica Investigator 
Award 2017-21 (HTW) contract number AS-IA-106-
M02 (HTW), the Kenda Foundation (JWC) 
as well as Contracts No. 104-2112-M-001-038-MY3 (HTW), 
No. 105-2112-M-002-017-MY3 (JWC), No. 104-2112-M-001-038-MY3 (CPL) 
from the Ministry of Science and Technology, and 
2017-18/ECP-2 from the National Center of Theoretical Physics Taiwan.

\bibliographystyle{apsrev4-1}
\bibliography{reference}

\end{document}